\documentclass[preprint,showpacs,preprintnumbers,amsmath,amssymb]{revtex4}

\usepackage{graphicx}

\begin{document}


\title{Cavity enhanced spin measurement of the ground state spin of an NV center in diamond}

\author{A.~Young}\email{A.Young@bristol.ac.uk}
\author{C.Y.~Hu}
\author{L. Marseglia}
\author{J.P. Harrison}
\author{J.L. O'Brien}
\author{J.G.~Rarity}
\address{Department of Electrical and Electronic Engineering,
University of Bristol, University Walk, Bristol BS8 1TR, United
Kingdom}

\begin{abstract}

We propose a high efficiency high fidelity measurement of the ground
state spin of a single NV center in diamond, using the effects of
cavity quantum electrodynamics. The scheme we propose is based in
the one dimensional atom or Purcell regime, removing the need for
high Q cavities that are challenging to fabricate. The ground state
of the NV center consists of three spin levels $^{3}A_{(m=0)}$ and
$^{3}A_{(m=\pm1)}$ (the $\pm1$ states are near degenerate in zero
field). These two states can undergo transitions to the excited
($^{3}E$) state, with an energy difference of $\approx6-10$ $\mu$eV
between the two. By choosing the correct Q factor, this small
detuning between the two transitions results in a dramatic change in
the intensity of reflected light. We show the change in reflected
intensity can allow us to read out the ground state spin using a low
intensity laser with an error rate of $\approx7\times10^{-3}$, when
realistic cavity and experimental parameters are considered. Since
very low levels of light are used to probe the state of the spin we
limit the number of florescence cycles, thereby limiting the non
spin preserving transitions through the intermediate singlet state
$^{1}A$.

\end{abstract}
\date{\today}


\maketitle

Addressing single spins is an important route to quantum
computation\cite{nielsen}. The long decoherence times of spins such
as trapped atoms\cite{cqed,langer:060502},
ions\cite{ci-prl-95-4091}, or charged quantum
dots\cite{PhysRevA.68.012310}, make them ideal candidates for
storing and processing quantum information. There are many schemes
for using internal spin states in all these
architectures\cite{PhysRevLett.92.127902,ci-prl-95-4091,lo-pra-98-120},
resulting in the demonstration of fundamental quantum logic
gates\cite{leibfr-nat-422-412,schmidt-nat-422-408,XiaoqinLi08082003}.
NV centers in diamond have long decoherence times even at room
temperature\cite{PhysRevLett.92.076401} making them another
promising candidate for performing quantum information tasks.
Several experiments have shown the manipulation of the ground state
spin of a diamond NV center using optically detected magnetic
resonance techniques
(ODMR)\cite{PhysRevB.64.041201,PhysRevLett.92.076401}. This has
further led to the coherent control of single $^{13}C$ nuclear spins
and quantum logic
operations\cite{PhysRevLett.93.130501,L.Childress10132006}. The main
problem in using ODMR is that the detection step involves observing
fluorescence cycles from the NV center, which has a probability of
destroying the spin memory. Since the energy level transitions of
the NV center are not polarization sensitive, we cannot use Faraday
rotations to perform quantum non demolition measurements of the spin
state as was shown for charged quantum
dots\cite{MeteAtature04282006,J.Berezovsky12222006}. The scheme we
propose here is similar to the ODMR scheme, however by the
introduction of a low Q cavity we vastly reduce the number of
photons required to probe the spin state, therefore keeping the
disturbance of the ground state spin to a minimum and not destroying
the spin memory.

\begin{figure}\label{fig:nvenergy}
\center
\includegraphics[width=10cm, height=7cm]{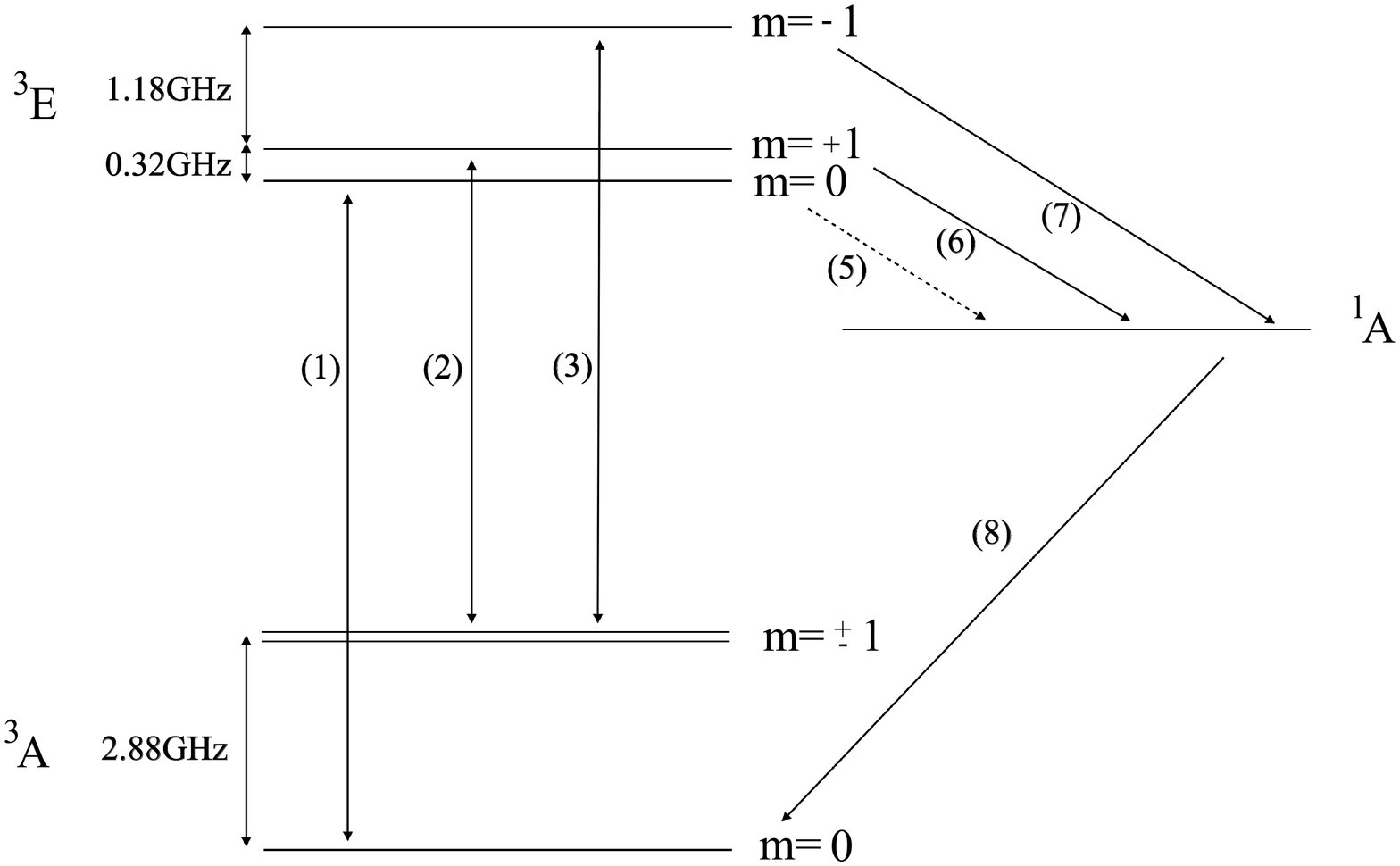}
\caption{Energy level diagram of the NV center in diamond showing
the experimentally determined ground and excited state
splitting\cite{fedor,loubser}. The defect has zero phonon line at
$637$ nm, with width of order MHz at low
temperatures\cite{0953-8984-18-21-S08}}
\end{figure}

If we consider the energy level structure of the NV center in figure
1, the ground state is a spin triplet split by $2.88$ GHz due to
spin-spin interactions\cite{loubser}. The excited state is a triplet
split by spin-spin interactions, but with the further addition of
spin-orbit coupling\cite{manson:104303}. Recent experimental
evidence\cite{fedor} has uncovered this excited state structure
(figure 1). The net effect of spin-spin and spin-orbit interactions
is to create a detuning $\approx1.4$ GHz($6$ $\mu$eV) between the
transition from the $^{3}A_{(m=0)}$ and the $^{3}A_{(m=+1)}$ to
$^{3}E$ state or $\approx2.5$ GHz($10$ $\mu$eV) for the
$^{3}A_{(m=-1)}$ It is exactly this detuning that we plan to exploit
to measure the ground state spin of the defect. The energy level
structure is not simply a ground and excited triplet state, there
also exists an intermediate singlet state $^{1}A$. There is a
probability that the $^{3}E$ state can decay to this state, with
different rates depending on the spin state. For the
$^{3}E_{m=\pm1}$ states (transitions 6,7) both theoretical
predictions and experimental results suggest the decay rate is
around $0.4\times
1/\tau$\cite{manson:104303,0953-8984-16-30-R03,PhysRevLett.85.290},
where $\tau$ is the spontaneous emission lifetime ($\approx13$ ns).
For the $^{3}E_{m=0}$ state (transition 5) theoretically the rate of
decay to the singlet should be zero\cite{manson:104303}, however
experimental observations have shown the rate to be
$\approx10^{-4}\times1/\tau$\cite{0953-8984-16-30-R03}. Since the
$^{1}A$ singlet state decays preferentially to the $^{3}E_{m=0}$
state\cite{manson-joflum-127-98,manson:104303} (transition 8), then
it is clear from the rates above that broadband excitation leads to
spin polarization in the spin zero ground state\cite{Santori:06}.
Since transition $8$ is non radiative then there will be a dark
period in the fluorescence when it becomes populated, and as the
decay rate from $^{3}E_{m=\pm1}$ to the singlet state is much larger
than from $^{3}E_{m=0}$, the change in intensity measures the spin
state\cite{0953-8984-16-30-R03}. Clearly using fluorescence
intensity to detect the spin state has a probability to flip the
spin, therefore it would seem necessary for a scheme to suppress
this. However spin flip transitions are essential to initialize the
system. Thus a compromise is required between the perfectly cyclic
spin preserving transitions required for readout, and the $\Lambda$
type spin flip transition required for initialization.

\begin{figure}\label{fig:schematic}
\center
\includegraphics[width=9.5cm, height=7cm]{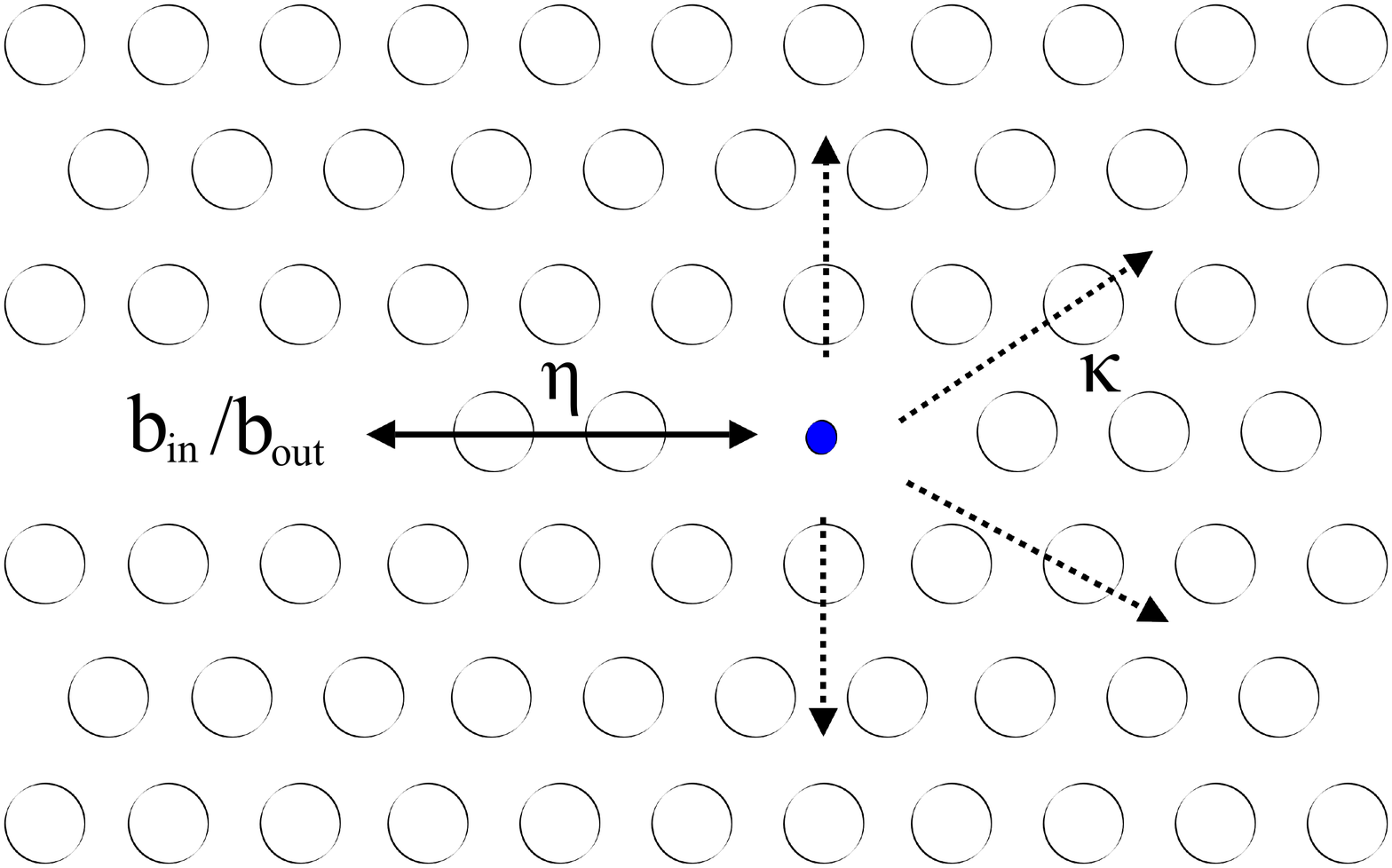}
\caption{Schematic diagram of an NV center embedded in a photonic
crystal cavity with cavity decay rate $\kappa$ coupled to a photonic
crystal waveguide at a rate $\eta$.}
\end{figure}

We consider the structure in figure 2, which can be modeled as a
single sided cavity, where $\kappa$ is the cavity decay rate (side
leakage), $\eta$ is the coupling of the cavity to external modes,
$g$ is the NV center cavity coupling rate and $\gamma$ is the NV
dipole decay rate. We can write down the Heisenberg equations of
motion for this structure as\cite{walls-milburn}:

\begin{eqnarray}
\frac{d\hat{a}}{dt}=-[i(\omega_{c}-\omega)+\frac{\eta}{2}+\frac{\kappa}{2}]\hat{a}-g\sigma_{-}-\sqrt{\eta}b_{in}\\
\frac{d\sigma_{-}}{dt}=-[i(\omega_{a}-\omega)+\frac{\gamma}{2}]\sigma_{-}-g\hat{\sigma_{z}}\hat{a}\\
\frac{d\sigma_{z}}{dt}=\gamma(1+\sigma_{z})-2g(\sigma_{-}\hat{a}^{\dag}+\hat{a}\sigma_{+})
\end{eqnarray}
where $\omega_{a}$ and $\omega_{c}$ are the atomic transition
($\sigma_{-}$) and the intracavity photon (annihilation operator
$\hat{a}$) frequencies respectively. $\hat{\sigma}_{z}$ represents a
Pauli Z operator on the atomic state and measures the population
inversion. If we now combine this with the input output relation for
this cavity:

\begin{equation}
b_{in}-b_{out}=\sqrt{\eta}a,
\end{equation}
then we can find the reflection coefficient for light input into the
cavity via $b_{in}$:

\begin{equation}
r(\omega)=\frac{b_{out}}{b_{in}}=\frac{[i(\omega_{a}-\omega)+\frac{\gamma}{2}][i(\omega_{c}-\omega)+\frac{\kappa}{2}-\frac{\eta}{2}]+g^2}{[i(\omega_{a}-\omega)+\frac{\gamma}{2}][i(\omega_{c}-\omega)+\frac{\kappa}{2}+\frac{\eta}{2}]+g^2}
\end{equation}
where we have set $\sigma_{z}=-1$ as is appropriate for the weak
excitation limit. At low temperature the zero phonon linewidth is
$0.1$ $\mu$eV\cite{0953-8984-18-21-S08}, we set $g=0.03$ meV as
appropriate for a cavity mode volume of $0.02$ $\mu$m$^{3}$, where
the NV has an oscillator strength of $\approx0.2$ given a $13$ ns
lifetime. It is desirable for the cavity to be critically coupled to
the input output so we will set $\eta$ to be $50$ times faster than
$\kappa$.

\begin{figure}\label{fig:refspec}\center
\includegraphics[width=13cm, height=9cm]{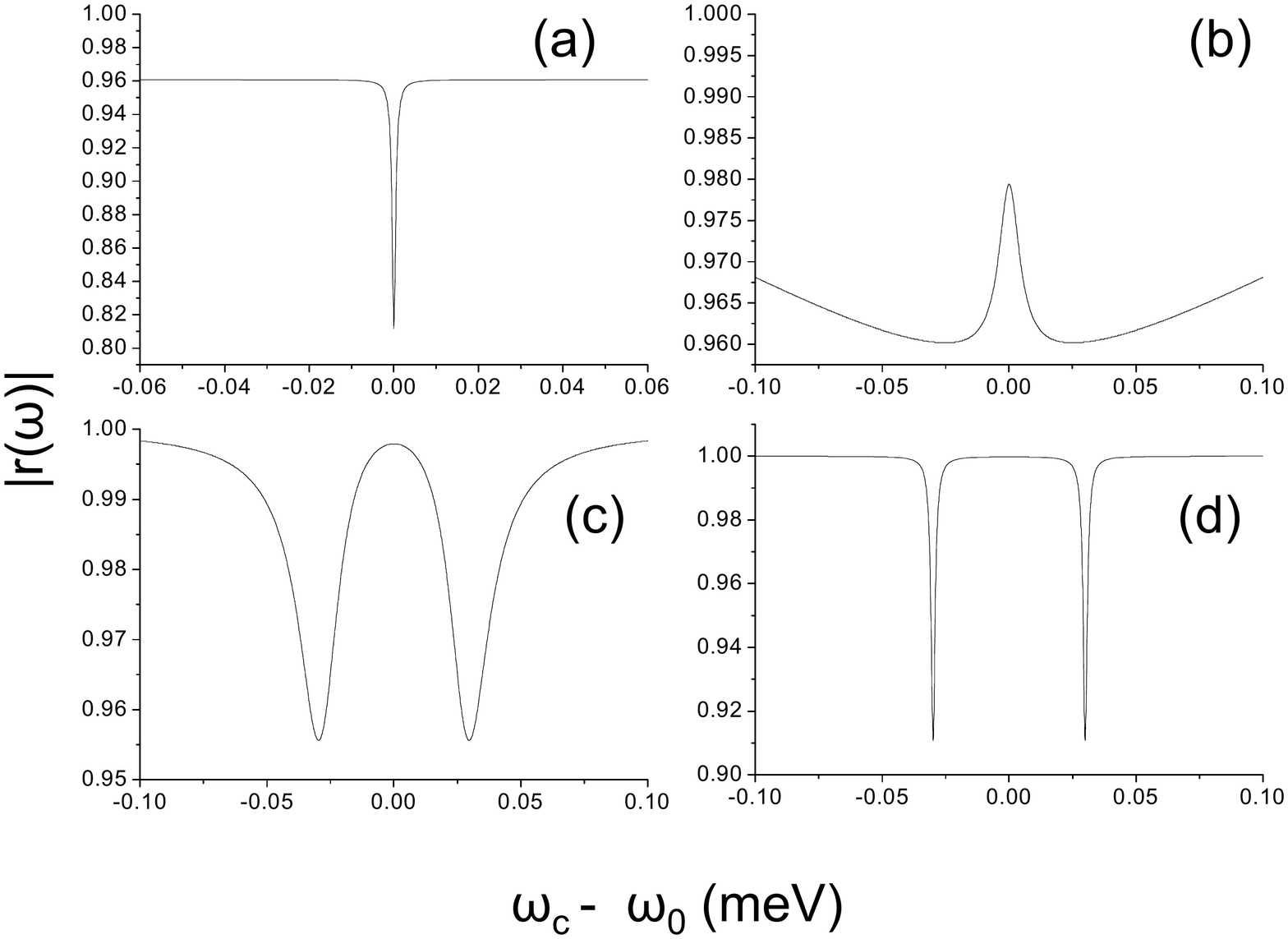}
\caption{Plots showing the reflection spectra against detuning where
$\omega_{a}=\omega_{c}$ for (a) $\kappa=75$ $\mu$eV, (b)$\kappa=7.5$
$\mu$eV,(c)$\kappa=0.75$ $\mu$eV,(d)$\kappa=0.075$ $\mu$eV, and
$\eta=50\kappa$ for all of the above}
\end{figure}

Figure 3 shows the effect of varying $\kappa$ and $\eta$ on the
reflection coefficient. When $\kappa$ and $\eta$ are very low we are
in the strong coupling regime with $g>>\kappa,\eta,\gamma$, here we
can clearly see the two Rabi split dressed atom-cavity states in
figure 3(d), and all of the light resonant with the cavity mode is
reflected. As we increase $\kappa$ and $\eta$ we are no longer able
to resolve the two states and cross over into the one dimensional
atom regime $\eta+\kappa>g>\gamma$, where in figure 3(b) there is a
small peak in reflectivity on resonance, a result of quantum
interference\cite{ji-it-pra-77-015806}. In this one dimensional atom
or Purcell regime the damping of the atomic transition (zero phonon
line) plays an increasingly dissipative role as a larger proportion
of the radiative decay is into non cavity modes. The result of this
is a dip in reflectance clearly visible in figure 3(a).

\begin{figure}\label{fig:refveta}\center
\includegraphics[width=10cm,height=7cm]{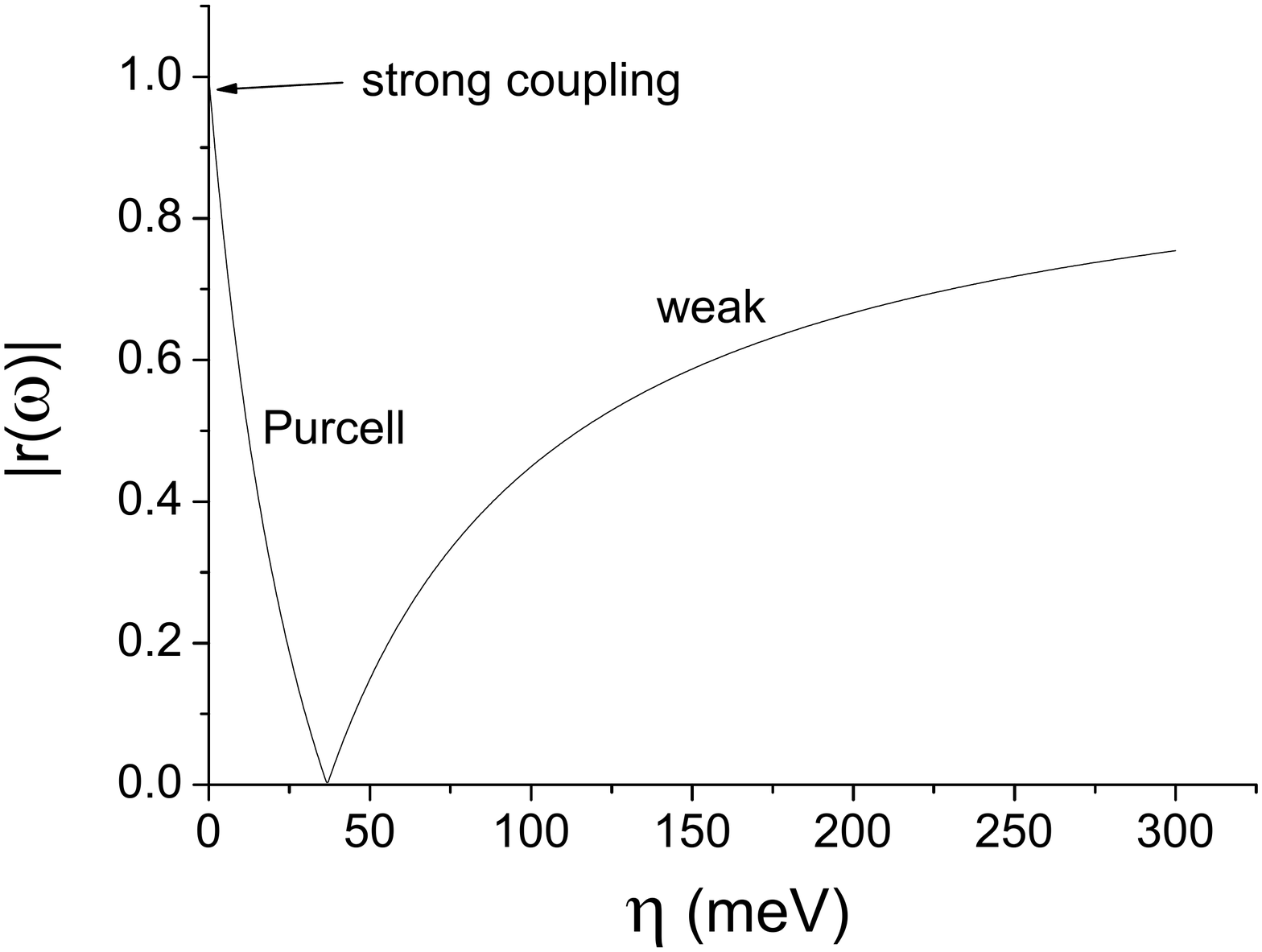}
\caption{Calculated reflectance against cavity damping at
$\omega_{c}=\omega_{a}$ and $\omega_{c}-\omega=0$, where we have
ignored $\kappa$ as $\eta=50\kappa$. The relevant atom-cavity
coupling regimes are labeled, the strong coupling regime occurs when
$\eta$ is very low and $|r(\omega)|\approx1$. The crossover from the
Purcell regime to the weak coupling regime occurs at a value of
$36$meV as predicted.}
\end{figure}

In figure 4 we have plotted the reflectance at the cavity-atom
resonance against $\eta$ as it is the dominant decay channel for the
cavity mode. The amount of reflected light drops to zero at a value
of $\eta=4g^{2}/\gamma$ which corresponds to the transition from the
one dimensional atom to the weak coupling regime. Thus at this point
all of the light absorbed by the NV center is emitted into non
cavity modes. After this turn over point we are in the weak coupling
regime, where the NV center has progressively less effect on the
dynamics of the system as it so weakly coupled. It is near this
transition region that we wish to operate where the narrow feature
in the reflection spectrum caused by the zero phonon linewidth
dominates (figure 5).

\begin{figure}\label{fig:refmain}\center
\includegraphics[width=10cm,height=7cm]{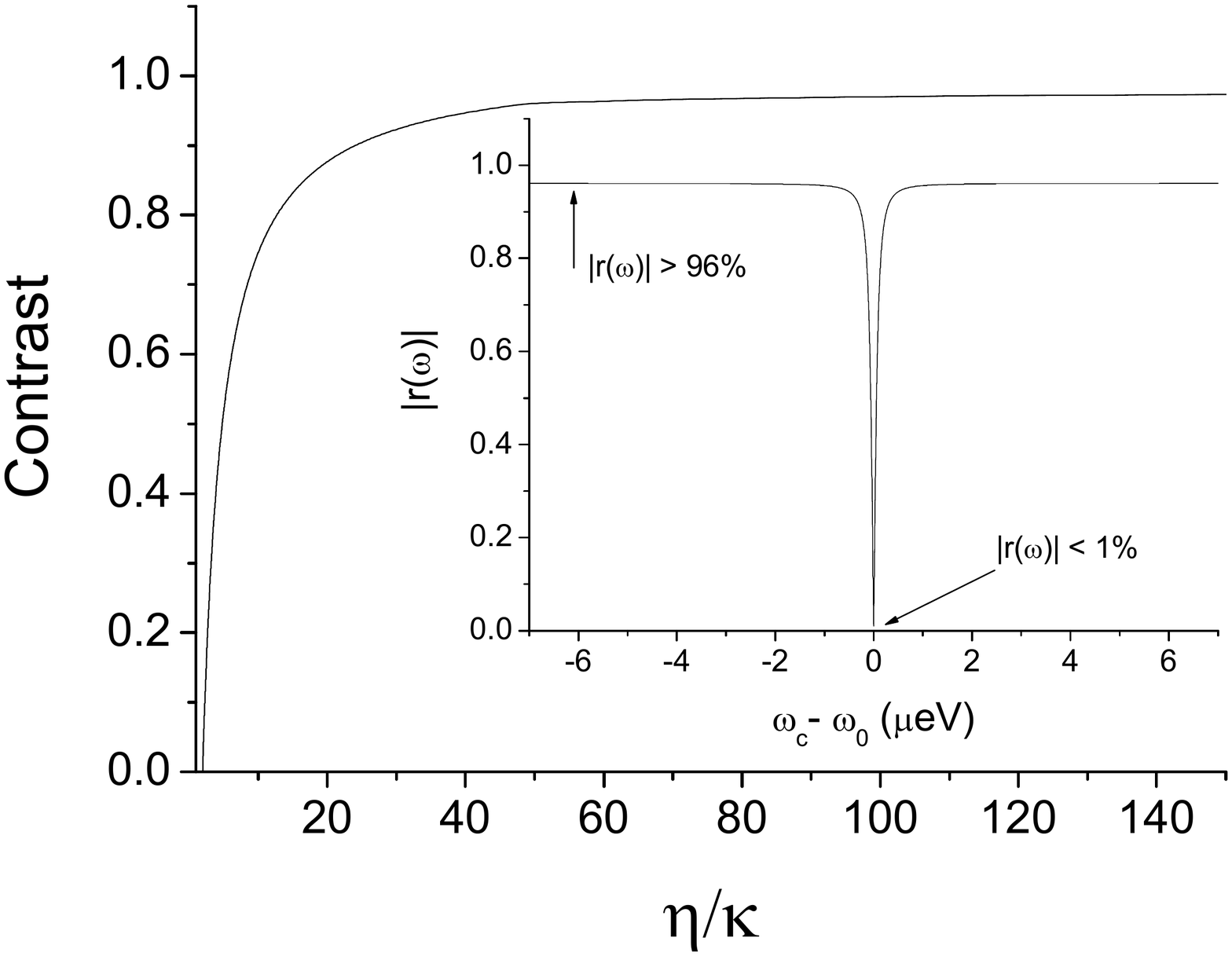}
\caption{Plot showing how intensity contrast
($|r(\omega)|^2_{m=\pm1}-|r(\omega)|^2_{m=0}$) varies with the ratio
of $\eta$ to $\kappa$. Inset: The reflection spectra against
detuning where $\omega_{a}=\omega_{c}$ for $\kappa=750$ $\mu$eV, and
$\eta=50\kappa$.}
\end{figure}

If we consider figure 5, then if we set the cavity to be resonant
with the $^{3}A_{(m=0)}$ to excited state transition then the
reflected intensity for resonant exitation becomes:

\begin{equation}
|r(\omega)|^2_{m=0}=\left|\frac{\gamma(\kappa-\eta)+4g^{2}}{\gamma(\kappa+\eta)+4g^{2}}\right|^2=0.001
\end{equation}
Almost none of the input light will be reflected when the NV center
is in the spin $m=0$ ground state. However if it is in the spin
$m=+1$($m=-1$) state then the $7$ $\mu$eV($10$ $\mu$eV) detuning
means that the NV center is effectively uncoupled giving a reflected
intensity of:

\begin{equation}
|r(\omega)|^2_{m=\pm1}=\left|\frac{(\kappa-\eta)}{(\kappa+\eta)}\right|^2=0.92
\end{equation}
Nearly all of the input light will be reflected. This contrast in
intensity ($|r(\omega)|_{m=\pm1}-|r(\omega)|_{m=0}$) can be easily
detected. What makes this result significant is that the curve in
figure 5 corresponds to a total Q factor
$Q_{tot}=\omega/(\kappa+\eta)\approx 55$. Since we have set $\eta$
to be $50$ times greater than $\kappa$, this means the photonic
crystal cavity before coupling needs to have
$Q=\omega/\kappa\approx3000$. This is much lower than the cavity Q
factor of $300000$ that would be required to have the cavity
linewidth narrow enough to resolve the two transitions, which when
coupled to a waveguide in the same way as here would need to exceed
$10^{7}$. It is possible to further reduce the requirements on the
cavity Q factor by reducing the the ratio of $\eta$ to $\kappa$.
However the result of this is a reduction in the intensity contrast
between the two spin states as a larger proportion of the light
confined in the cavity leaks out of the side. The intensity contrast
which measures the spin is not influenced by total Q factor, the
optimal value being $Q_{tot}\approx55$, the contrast is only
influenced by the ratio of $\eta$ to $\kappa$. It is desirable to
have this contrast at a maximum in order to minimize errors in state
identification.

There are several benefits to this scheme. The first is the obvious
increase in collection efficiency of the photons, making low
intensity measurements possible. Since less photons are required to
probe the spin state there are less fluorescence cycles therefore a
reduced probability of a spin flip transition. Additionally as the
cavity is resonant with the $m=0$ transition by probing with narrow
band light then we never excite the spin $\pm1$ transitions which
have a higher probability to spin-flip, hence the system is
optimized for spin preserving transitions. However if we pump with a
broad band laser source we can easily spin polarize the ground state
to initialize the system. Since we are in the low Q regime then the
Purcell factor is small, $F_{p}\approx4$ for a system with the
parameters listed above, thus the rate of spontaneous emission(SE)
into the cavity is not significantly modified. If we were operating
in the high Q or strong coupling regime then the SE rate into the
cavity would be much larger, and the decay rate to the $^{1}A$
singlet state would remain unmodified. Therefore with the
probability of a spin flip transition greatly reduced, the system
would not be simultaneously optimized for readout and
initialization.

In order to make the scheme experimentally relevant the limitations
of current detector technology must be included. If we consider an
overall detection efficiency of $33\%$ then if we input $60$ photons
we can expect to detect $20$ with unit reflectivity. If the spin is
in the $m=\pm1$ state ($|r(\omega)|^2\approx92\%$) it is reasonable
to expect $18$ photons to be detected. If the spin is in the $m=0$
state ($|r(\omega)|^2<1\%$) we may expect $1$ photon to be detected.
If we set a detection threshold of $6$ photons the error in the
measurement can then be found from the probability of detecting $>6$
photons when we expect $1$ and the probability of detecting $<6$
photons when we expect $18$, giving an error rate of
$\approx1.5\times10^{-3}$ (assuming poissonian distribution).
Standard silicon avalanche photo diodes have a dead time of $50$ ns
which means that it will take $3$ $\mu$s to carry out a measurement
with $18$ detected photons(running at one third of the detector
saturation count rate). Since the longest observed spin coherence
time of an NV center is $600$ $\mu$s\cite{P.Neumann06062008} this
introduces a further error rate of $5.5\times10^{-3}$. There are
also errors associated with saturation of the NV center. However as
the detector dead time is much larger than the modified spontaneous
emission lifetime then these are negligible. Finally there is also
an error associated with decay from the $^{3}E_{m=0}$ to the
$^{3}E_{m=\pm1}$ state via the $^{1}A$ singlet state which for $60$
photons is $<10^{-3}$. Thus the total error rate is
$\approx7\times10^{-3}$.

Simulations of photonic crystals in diamond have shown Q factors
larger than $10^6$ for mode volumes around $0.02$ $\mu$m$^{3}$, or
larger than $10^{5}$ for mode volumes around $0.008$
$\mu$m$^{3}$\cite{Zhang:04}. These values are significantly more
demanding than those required for this scheme, particularly the Q
factor. Experimental evidence suggests that the actual Q factors
will be much lower than those simulated. Cavities fabricated by Wang
et.al.\cite{wang:201112}, showed more than a factor of $10$
shortfall in the experimental Q factor compared to the simulated,
attributed to defects in the nanocrystalline structure. Nevertheless
their measured Q factor of $585$ would allow a ratio of
$\eta\approx10\kappa$, in order to have an overall Q factor of $55$.
This would result in a $65\%$ contrast between the two spin states,
increasing the error rate to $\approx2\times10^{-2}$. Theoretical
considerations of the absorption in nanocrystalline diamond have
predicted a reduction in Q factor from a value of $66300$ to a value
of around $1350$ for a cavity of mode volume $0.02$
$\mu$m$^{3}$\cite{Kreuzer:08}. For our purposes this would result in
a contrast of $85\%$, where the error rate would be
$\approx1\times10^{-2}$. So the scheme is clearly robust and can
cope with experimental imperfections. The main difficulty with this
scheme, which is true for all schemes, is the positioning of the NV
center at the field maximum. If the precision is poor then this can
have a detrimental effect on the coupling rate $g$. This in turn
reduces the intensity contrast between the two spin states, which is
sensitive to the value of $g$ compared to the zero phonon linewidth
$\gamma$. There is promise that ion implantation in single crystal
diamond could hold the key to fabricating suitable
devices\cite{0953-8984-18-21-S09}, the precision of implantation is
currently on the nanometer
scale\cite{0953-8984-18-21-S09,nat_phs_2_408}. The use of single
crystal diamond would dramatically reduce the absorption losses
caused by defects, so experimental Q factors should be closer to the
theoretical predictions.

In conclusion we have proposed an efficient low error measurement of
the ground state spin of an NV center. For the realistic parameters
proposed here we can achieve error rates of around $7\times10^{-3}$.
The setup can easily switch between initialization and readout by
switching from a broad to narrow band laser source. Low error
readout requires modest Q factors, and even with current limitations
of photonic crystal cavities the error rate could be as low as
$2\times10^{-2}$. Work needs to be done on the design and
fabrication of photonic crystal cavities coupled to waveguides,
particularly in single crystal diamond to minimize absorption
losses. We also note that we can measure the spin with a single
photon with $92\%$ fidelity (assuming ideal detection), where
fidelity is simply the contrast between the two spin states. Hence
with some modifications the ideas here could be used to remotely
entangle two spatially separated NV centers embedded in cavities,
which is a subject for further study.

We acknowledge support from the UK EPSRC (QIP IRC), and the European
Commission under projects IST-015848-QAP and IST-034368 EQUIND and
Nanoscience ERA project NEDQIT. J.G.R. is supported by a Royal
society Wolfson Merit award.

\end{document}